\documentclass[journal,twoside]{IEEEtran}
\usepackage{indentfirst}
\usepackage{amsfonts}
\usepackage{amsmath}
\usepackage{amsthm}
\usepackage{graphicx}
\usepackage{fancyhdr}
\usepackage{float}
\usepackage{commath}
\usepackage[utf8]{inputenc}
\usepackage{amssymb}
\usepackage{enumerate}
\usepackage[shortlabels]{enumitem}
\usepackage[dvipsnames]{xcolor}
\usepackage{soul}
\usepackage{listings}
\usepackage[hidelinks,colorlinks=true,linkcolor=blue,citecolor=blue,urlcolor=black]{hyperref}
\usepackage{bm}
\usepackage{cite}
\usepackage{flushend}
\usepackage{lipsum}
\usepackage{adjustbox}
\usepackage{tabularx}
\usepackage{tikz}
\usepackage{pgfplots}
\usetikzlibrary{patterns}
\usepgfplotslibrary{fillbetween,colorbrewer}
\usepgfplotslibrary{groupplots}
\usetikzlibrary{positioning,shadows,shapes,backgrounds,calc,spy,arrows}
\usetikzlibrary{intersections}
\usepackage{pgfplotstable}
\pgfplotsset{compat=newest}
\usepackage[nolist]{acronym}
\usepackage{booktabs}
\usepackage{threeparttable}
\usepackage{tcolorbox}
\renewcommand{\arraystretch}{1.5}
\usepackage{caption}
\theoremstyle{definition}
\newtheorem{example}{Example}

\definecolor{codegreen}{rgb}{0,0.6,0}
\definecolor{codegray}{rgb}{0.5,0.5,0.5}
\definecolor{codepurple}{rgb}{0.58,0,0.82}
\definecolor{backcolour}{rgb}{0.99,0.99,0.99}

\definecolor{QAMcross}{RGB}{0,154,86}
\definecolor{QAMfcross}{RGB}{69,99,168}
\definecolor{QAMecoc}{RGB}{161,33,33}
\definecolor{QAMgs}{RGB}{152,78,163}
\definecolor{QAMub}{RGB}{128,128,128}

\newcommand{\Xset}{\mathcal{X}}
\newcommand{\XkO}{\mathcal{X}_k^0}
\newcommand{\XkI}{\mathcal{X}_k^1}

\newcommand{\ER}{\mathrm{ER}}

\newcommand{\Lawgn}{L_k^\mathrm{awgn}}
\newcommand{\Lzc}{L_k^\mathrm{zc}}

\definecolor{colorLk}{RGB}{255,0,0}
\definecolor{colorLawgn}{RGB}{0,0,255}
\definecolor{colorLzc}{RGB}{0,0,0}

\definecolor{greenALT}{RGB}{0,154,86}
\definecolor{blueALT}{RGB}{53,133,187}
\definecolor{redALT}{RGB}{227,26,27}

\DeclareCaptionFont{smallcaps}{\scshape}  
\begin{document}
\begin{acronym}[]
    \acro{IM}{intensity-modulation}
    \acro{DD}{direct-detection}
    \acro{SNR}{signal-to-noise ratio}
    \acro{RIN}{relative intensity noise}
    \acro{RV}{random variable}
    \acro{SER}{symbol error rate}
    \acro{PAM}{pulse amplitude modulation}
    \acro{QAM}{quadrature amplitude modulation}
    \acro{PDF}{probability density function}
    \acro{PMF}{probability mass function}
    \acro{PS}{probabilistic shaping}
    \acro{GS}{geometric shaping}
    \acro{ES}{equally-spaced}
    \acro{FEC}{forward error correction}
    \acro{BER}{bit error rate}
    \acro{MAP}{maximum a-posteriori probability}
    \acro{ML}{maximum likelihood}
    \acro{AWGN}{additive white Gaussian noise}
    \acro{OMA}{optical modulation amplitude}
    \acro{ER}{extinction ratio}
    \acro{RC}{raised-cosine}
    \acro{RRC}{root-raised-cosine}
    \acro{MZM}{Mach-Zehnder modulator}
    \acro{SMF}{single-mode fiber}
    \acro{ISI}{intersymbol interference}
    \acro{PD}{photodiode}
    \acro{TIA}{transimpedance amplifier}
    \acro{ADC}{analog-to-digital converter}
    \acro{DAC}{digital-to-analog converter}
    \acro{PSD}{power spectral density}
    \acro{DSP}{digital signal processing}
    \acro{DCI}{datacenter interconnect}
    \acro{P/S}{parallel-to-serial}
    \acro{S/P}{serial-to-parallel}
    \acro{HD}{hard-decision}
    \acro{SD}{soft-decision}
    \acro{BICM}{bit-interleaved coded modulation}
    \acro{BW}{bit-wise}
    \acro{SW}{symbol-wise}
    \acro{AIR}{achievable information rate}
    \acro{AIRs}{achievable information rates}
    \acro{BSC}{binary symmetric channel}
    \acro{RS}{Reed-Solomon}
    \acro{MI}{mutual information}
    \acro{GMI}{generalized mutual information}
    \acro{NGMI}{normalized GMI}
    \acro{LLR}{log-likelihood ratios}
    \acro{OH}{overhead}
    \acro{LRB}{least reliable bits}
    \acro{FFE}{feedforward equalization}
    \acro{DFE}{decision feedback equalization}
    \acro{LDPC}{low-density parity-check}
\end{acronym}

\title{Low-complexity Soft-decision LLR Calculations for Next-generation IM-DD Systems with RIN}

\author{Felipe~Villenas,~\IEEEmembership{Graduate Student Member,~IEEE,} Yunus~Can~G\"{u}ltekin,~\IEEEmembership{Member,~IEEE,} and Alex~Alvarado,~\IEEEmembership{Senior Member,~IEEE}
\thanks{This research is part of the project COmplexity-COnstrained LIght-coherent optical links (COCOLI) funded by Holland High Tech $|$ TKI HSTM via the PPS allowance scheme for public-private partnerships. \textit{(Corresponding author: Felipe Villenas.)}}%
\thanks{The authors are with the Information and Communication Theory Lab, Signal Processing Systems Group, Department of Electrical Engineering, Eindhoven University of Technology, 5600 MB Eindhoven, the Netherlands (e-mails: \{f.i.villenas.cortez, y.c.g.gultekin, a.alvarado\}@tue.nl).}%
}

\markboth{Pre-print}
{Villenas \MakeLowercase{\textit{et al.}}: Low-Complexity Soft-Decision LLR Calculations for Next-Generation IM-DD Systems with RIN}



\maketitle

\begin{abstract}
The demand for higher speeds in intra-data center interconnects will eventually require high-order pulse amplitude modulation (PAM) combined with soft-decision (SD) forward error correction (FEC). The laser relative intensity noise (RIN) is an important noise impairment that limits the performance of high-speed intensity-modulation (IM) and direct-detection (DD) systems, as it induces a channel with signal-dependent noise. In this paper, we show that an accurate calculation of the log-likelihood ratios (LLRs) is critical for the performance of the SD-FEC decoder in RIN-dominated IM-DD systems. First, we show that assuming signal-independent additive white Gaussian noise (AWGN) statistics to compute LLRs results in a significant penalty in performance. As an alternative, we propose a low-complexity piecewise linear approximation of the exact LLRs of Gray-coded PAM-4 and PAM-8 for the RIN-dominated IM-DD channel. We show using the generalized mutual information that our approximation results in no performance loss versus using exact LLRs, unlike the AWGN-like assumption. Furthermore, we validate the analysis with bit error rate (BER) performance of SD decoding. We show that our approximation matches the BERs achieved using exact LLRs. Therefore, our approximation avoids a BER error floor penalty of up to 7.2 times using an extended Hamming code with PAM-4, and up to 2.41 dB of optical modulation amplitude penalty using high-code-rate low-density parity-check codes with PAM-8 for the same target BER.
\end{abstract}


\section{Introduction}

\begin{figure*}[!t]
    \centering
    \includegraphics[]{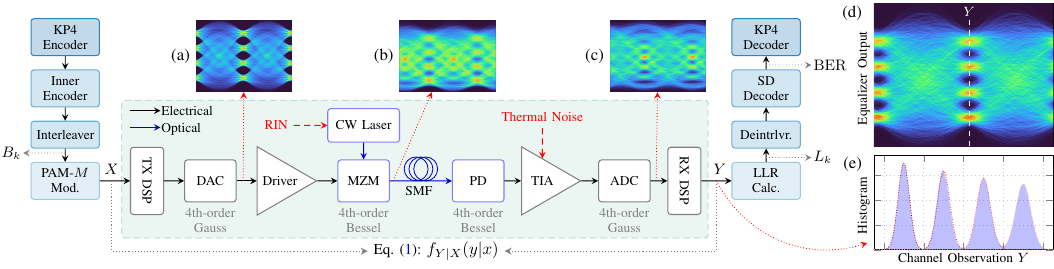}
    \caption{Considered system: concatenated KP4+SD-Inner FEC scheme with an IM-DD channel with signal-dependent noise. The insets show the eye diagram of a 226 GBd transmitted PAM-4 signal through a MZM/PD with a 3-dB bandwidth of 120 GHz, and a DAC/ADC with 110 GHz. The insets show the eye diagram at the output of the (a) DAC, (b) MZM, (c) ADC, and (d) equalizer. The inset (d) and histogram in (e) show that each PAM level has a different noise distribution due to RIN. The output $Y$ is then used to calculate the LLRs $L_k$ with either \eqref{eq:Lk_2}, \eqref{eq:Lk_awgn}, or \eqref{eq:L1_zc}$-$\eqref{eq:L3_zc_pam8}, and then fed to the SD decoder.}
    \label{fig:system_model}
\end{figure*}

\IEEEPARstart{T}{he} rapid growth of artificial intelligence applications is driving demand for higher speeds in short-reach data center interconnects (DCI) \cite{siamak2026}. Low hardware costs and low power consumption are essential, and thus, optical transceivers for DCI employ intensity-modulation (IM) and direct-detection (DD) with $M$-ary pulse amplitude modulation (PAM-$M$) \cite{zhou2026}. Current systems employ PAM-4 at $200$ Gb/s/lane. However, scaling the data rates to 400 Gb/s/lane and beyond faces significant challenges due to the bandwidth constraints of electro-optical components and noise impairments \cite{pang2020200}. An important noise impairment in such systems is the relative intensity noise (RIN) arising from the laser source \cite{tatarczak2026}.

To achieve the scaling to 400 Gb/s/lane and beyond, several technologies have been studied, such as advanced \ac{DSP} techniques \cite{oettinghaus2026advanced} or advanced \ac{FEC} codes\cite{wang2023advancedFEC}. The FEC standard for current DCI links is the \ac{HD} KP4 code \cite{baseline_200G}. To improve the coding gain with respect to KP4, several proposals have been made. For example, a concatenated FEC scheme was proposed for the 200 Gb/s/lane baseline in \cite{farhood_P802_3dj}, with implementation in commercial products such as \cite{marvell2024ara}. The concatenated FEC scheme consists of an outer \ac{HD} KP4 code, and an inner \ac{SD} extended Hamming code. The inner code aims to correct errors that come from optical propagation with a relatively simple FEC code (error correcting capability of $1$ bit) combined with \ac{SD} decoding, to increase the total coding gain. However, the demand for higher throughput will inevitably require an increase in modulation formats beyond PAM-4 combined with stronger FEC capabilities \cite[Sec.~II]{oettinghaus2026advanced}.

Soft information is used for \ac{SD} decoding, typically in the form of log-likelihood ratios (LLR). Calculating LLRs requires proper knowledge of the channel, which is not always straightforward to obtain accurately \cite{bosco2003soft}. Channels with signal-dependent noise, such as RIN, induce different noise statistics per symbol. LLR calculations for PAM systems with signal-dependent noise have been studied in the following works. The authors of \cite{shi2025enhanced} proposed a modified channel law to calculate the LLRs in a spatial division multiplexed IM-DD system, where the source of signal-dependent noise is optical amplification. In \cite{pan2023adaptive}, the authors proposed an adaptive LLR calculation algorithm for coherent modulation, where each symbol has its own noise variance, also due to optical amplification. Lastly, the authors of \cite{zhou2020improved} studied an IM-DD system with RIN from a directly modulated laser, and calculated the PAM-8 LLRs using a different noise variance for each symbol. 

To reduce the computational complexity of LLR calculations, demappers based on lookup tables precompute LLRs offline at the cost of storage \cite{wu2023low}, \cite[Ch. 3.3.3]{bicmBook}. Moreover, for the additive white Gaussian noise (AWGN) channel, LLRs are well approximated by piecewise linear functions of the received values at high signal-to-noise ratio \cite{ivanov2014asymptotic}. Furthermore, the number of pieces in the piecewise linear function can be further reduced by using local linearization of the LLRs \cite{tosato2002simplified,alvarado2009distribution}. In this paper, we adapt this LLR local-linearization approach to high-speed IM-DD systems with signal-dependent noise, such as RIN. To the best of our knowledge, LLR calculations for such systems have not been previously addressed using a low-complexity structure. We study three different LLR calculations for an IM-DD system limited by RIN. We use exact LLR calculation of this channel as the baseline and compare it with two low-complexity calculations based on mismatched LLRs. The first one, averaging the noise similar to approximate it as signal-independent AWGN, and the second, a piecewise linear approximation around the zero-crossing (ZC) points of the exact LLRs. The latter approximation has been shown to have negligible \ac{BER} performance loss after SD decoding in the AWGN channel \cite{tosato2002simplified}. Thus, we propose to extend it to the RIN-dominated case. 

We extend the work of our ECOC paper \cite{villenas2026ecoc} in which we showed that for this channel, \mbox{PAM-4} LLRs follow a quadratic polynomial relationship as a function of the received values. Furthermore, we showed that with our piecewise linear approximation of the LLRs for the RIN-dominated channel, there is virtually no post-FEC BER performance loss when using our proposed approximation as the soft information for an extended Hamming code with SD-FEC. In addition, in this paper we present the following extensions:
\begin{itemize}[label=\textbullet]
    \item First, we extend our developed approximation to PAM-8 and show the results for the extended Hamming code.
    \item We include a performance analysis of our approximation under estimation errors and bandwidth-limited scenarios.
    \item Then, we perform an analysis on \ac{AIRs} for the three different LLR calculations and analyze the impact of RIN on the AIRs.
    \item We show the post-FEC \ac{BER} performance for \ac{LDPC} codes, and link these results to the AIRs analysis.
\end{itemize}

The remainder of the paper is organized as follows. Sec.~\ref{sec:system} describes the IM-DD system model in detail. Sec.~\ref{sec:llr} details the proposed low-complexity LLR approximation and its robustness against estimation errors and residual ISI. Sec.~\ref{sec:gmi} studies the achievable information rates of the LLR calculations, and finally, the conclusions are given in Sec.~\ref{sec:conclusions}.
\section{System Model} \label{sec:system}

We consider a concatenated FEC scheme depicted in Fig.~\ref{fig:system_model}. The KP4-coded bits coming from the host are encoded with an inner code to generate the bits $\boldsymbol{B}=[B_1,\dotsc,B_k]$,\footnote{\textit{Notation Convention:} Capital letters (e.g., $X$) denote a \ac{RV}, while their lowercase version (e.g., $x$) denotes the RV realization. Bold letters (e.g., $\boldsymbol{X}$) are used to denote random vectors. The probability density function of a continuous RV $Y$ conditioned on $X$ is denoted by $f_{Y|X}(y|x)$, and the operator $\mathbb{E}_{X,Y}[\cdot]$ denotes the joint expectation with respect to $X$ and $Y$. Lastly, calligraphic letters (e.g., $\mathcal{X}$) denote sets.} where $k=1,2,\dotsc,m$ is the bit position, and $B_k\in\mathcal{B}=\{0,1\}$. The bits $\boldsymbol{B}$ are then interleaved and modulated into PAM-$M$ symbols $X\in\Xset$, with $M=2^m$, using the binary reflected Gray code (BRGC). In this paper, we use the constellations $\Xset=\{\pm\Delta,\pm3\Delta\}$ for PAM-4, and $\mathcal{X}=\{\pm\Delta,\pm3\Delta,\pm5\Delta,\pm7\Delta\}$ for PAM-8, where $\Delta$ is a scaling factor defined by the optical modulation amplitude (OMA) and link losses.
\subsection{Optical Channel}
The symbols $X$ are the input to the IM-DD system shown in Fig.~\ref{fig:system_model}. The transmitter (TX) DSP applies upsampling, pulse shaping, and amplitude clipping. The digital-to-analog converter (DAC) generates the analog waveform which is then conditioned by the driver for IM in the linear regime of a Mach-Zehnder modulator (MZM) in push-pull configuration, biased at its quadrature point. The frequency response of the DAC is modeled as a 4th-order Gauss filter, while the MZM is modeled as a 4th-order Bessel filter \cite{wettlin2026modulation}. The MZM is used to modulate the intensity of an O-band ($1310$ nm) continuous-wave (CW) laser with RIN. Here, we consider the RIN to have Gaussian statistics \cite[Sec.~3.3.6]{hui2019introduction} and a double-sided power spectral density (PSD) defined by the laser parameter in units of dB$/$Hz.

The optical signal is transmitted through a single mode fiber (SMF) of length $<\!500$ m. Since the signal is transmitted at the zero-dispersion wavelength, chromatic dispersion is negligible and the fiber attenuation is $0.35$ dB/km. At the receiver (RX), a photodiode (PD) generates a current proportional to the intensity of the incident optical light. The frequency response of the PD is also modeled as a 4th-order Bessel filter. The signal bias from IM is removed and a transimpedance amplifier (TIA) converts the current to voltage. During this process, thermal noise is generated, which is modeled as AWGN with double-sided PSD defined by the TIA input-referred noise (IRN) parameter in units of pA$/\sqrt{\mathrm{Hz}}$.

The analog-to-digital converter (ADC), with frequency response modeled as a 4th-order Gauss filter, samples the received analog waveform to be processed further by the receiver DSP. The RX DSP applies matched filtering, time alignment, equalization to remove intersymbol interference (ISI), and downsampling. The main challenge of current IM-DD systems targeting high speeds is to overcome the ISI due to bandwidth limitations. As symbol rate increases, so do the bandwidth requirements, and the DSP becomes more challenging \cite{oettinghaus2026advanced}. For this work, we consider that the ISI caused by bandwidth-limited components is removed using a $T/2$-spaced feed-forward equalizer (FFE). Sequence-based estimation methods, such as the soft-output Viterbi algorithm, are beyond the scope of this paper in order to maintain a low-complexity receiver DSP. The inset (d) of Fig.~\ref{fig:system_model} shows an example of the eye diagram of a PAM-4 signal after FFE. It can be seen that at the optimal sampling instant, each PAM level has a different noise distribution due to RIN. Furthermore, we consider that any residual ISI after FFE is negligible in order to isolate the effect of RIN. The inset (e) in Fig.~\ref{fig:system_model} shows a histogram of the channel observations $Y$ where the effect of RIN is clearly visible, resulting in a noise distribution proportional to the square of the symbol's optical power.

Given the description of the system model, we consider a memoryless decoding metric where the channel observation $Y$ only depends on the transmitted symbol $X$ in the current time instant via the \ac{PDF} \cite{villenas2026jstqe}
\begin{equation} \label{eq:cPDF}
    f_{Y|X}(y|x) = \frac{1}{\sqrt{2\pi\sigma^2(x)}}\mathrm{exp}\left(-\frac{(y-x)^2}{2\sigma^2(x)}\right).
\end{equation}
Although the channel might have memory from residual ISI, from now on we treat $f_{Y|X}(y|x)$ as the true channel transition probability. Note that \eqref{eq:cPDF} follows a Gaussian distribution but with a variance that depends on $x$ through the function $\sigma^2(x)$ defined as \cite[Eq.~(2)]{villenas2026jstqe}, \cite[Eq.~(10)]{szczerba20124}, \cite[Eq.~(8)]{rizzelli2023analytical} 
\begin{equation} \label{eq:noise_model}
    \sigma^2(X) \triangleq p_0 + p_1(X+\beta)^2,
\end{equation}
where $p_0$ is proportional to the square of the TIA IRN, while $p_1$ is proportional to the RIN parameter in linear scale. The IM bias \mbox{$\beta=(M-1)\Delta\frac{\ER+1}{\ER-1}\geq |\min\{\Xset\}|$} is modeled after the extinction ratio (ER) of the system \cite[Sec.~III]{che2021does}. The inset (e) of Fig.~\ref{fig:system_model} also shows the fit of the noise model \eqref{eq:noise_model} to the histogram data.

\subsection{Log-likelihood Ratio Calculations}
The PDF \eqref{eq:cPDF} is needed for the LLR calculation block (see Fig.~\ref{fig:system_model}). The resulting LLRs $\boldsymbol{L}=[L_1,L_2,\dotsc,L_k]$ are the soft-information fed to the SD decoder to decode the inner code after deinterleaving. After decoding, the decoded bits are passed to the KP4 decoder. In this paper, we analyze metrics up to the output of the SD decoder. Hence, we refer to BER as the post-FEC BER of the inner code.

For equally likely symbols $X$ and channel observation $Y$, the LLR for the bit position $B_k$ is calculated as
\begin{equation} \label{eq:L_initial}
    L_k = \log \frac{f_{Y|B_k}(y|1)}{f_{Y|B_k}(y|0)}=\log \frac{\sum_{x\in\XkI} f_{Y|X}(y|x)}{\sum_{x\in\XkO} f_{Y|X}(y|x)},
\end{equation}
where $\mathcal{X}_k^b\subset \Xset$ is the subset of symbols where the bit in position $k$ has a value of $b\in\mathcal{B}$. Note that $\eqref{eq:L_initial}$ requires knowledge of the true channel transition probability $f_{Y|X}(y|x)$, and thus from here on, we refer to $L_k$ as the \textit{exact} LLR. Using the channel PDF \eqref{eq:cPDF} in \eqref{eq:L_initial} results in
\begin{equation}
    L_k = \log \frac{\sum_{x\in\mathcal{X}_k^1} \frac{1}{\sigma(x)}\mathrm{exp}\left(-\frac{(y-x)^2}{2\sigma^2(x)}\right)}{\sum_{x\in\mathcal{X}_k^0} \frac{1}{\sigma(x)}\mathrm{exp}\left(-\frac{(y-x)^2}{2\sigma^2(x)}\right)}. \label{eq:Lk_2}
\end{equation}

Under the mismatched decoding framework, the receiver is optimized for an auxiliary channel with a decoding metric $q_{Y|X}(y|x)$ \cite{merhav1994information}. In this scenario, a \textit{mismatched} LLR \cite[Eq.~(4.76)]{bicmBook} is defined by replacing $f_{Y|X}(\cdot|\cdot)$ for $q_{Y|X}(\cdot|\cdot)$ in \eqref{eq:L_initial}. A receiver that is unaware of RIN would be optimized for an AWGN decoding metric $q_{Y|X}(y|x) = \mathrm{exp}\left(-\frac{(y-x)^2}{2\overline{\sigma}^2}\right)$, where the noise variance $\overline{\sigma}^2$ would be calculated as an average over all transmitted symbols. With this metric $q_{Y|X}(y|x)$, the mismatched LLRs become
\begin{align} \label{eq:Lk_awgn}
    \Lawgn &= \log \frac{\sum_{x\in\XkI} \mathrm{exp}\left(-\frac{(y-x)^2}{2\overline{\sigma}^2}\right)}{\sum_{x\in\XkO} \mathrm{exp}\left(-\frac{(y-x)^2}{2\overline{\sigma}^2}\right)}\\
    &\approx \frac{1}{2\overline{\sigma}^2}\left(\min_{x\in\XkO} (y-x)^2-\min_{x\in\XkI} (y-x)^2\right).\label{eq:L_maxlog_awgn}
\end{align}
This coincides with the LLR calculation of a linear AWGN channel, and hence we denote \eqref{eq:Lk_awgn} as $\Lawgn$. Most bit-wise decoders for IM-DD systems calculate LLRs under an AWGN channel assumption (see, e.g., \cite{yamamoto2025200}), as it is more straightforward than estimating the noise variance for each symbol. Furthermore, to reduce the computational complexity of $\eqref{eq:Lk_awgn}$, the well known max-log approximation \cite{viterbi1998intuitive} is used, which approximates $\Lawgn$ as \eqref{eq:L_maxlog_awgn} \cite[Eq.~(31)]{alvarado2015replacing}. Using $\Lawgn$ results in a clear mismatch with the described IM-DD system, as it completely neglects the signal-dependency of RIN.
\begin{figure}[!t]
    \centering
    \includegraphics[]{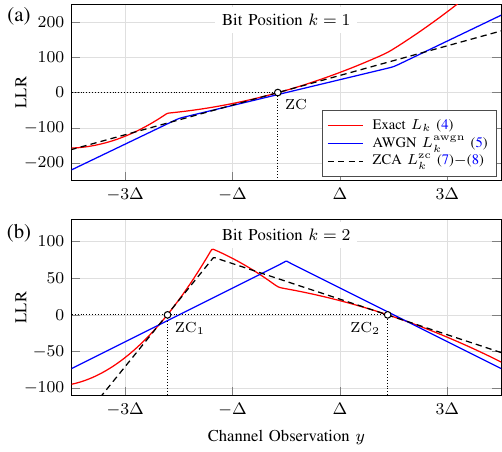}
    \caption{PAM-4 LLRs $L_k$, $\Lawgn$, and $\Lzc$ using parameters in Table~\ref{tab:sim_param} for bit position (a) $k=1$, and (b) $k=2$.}
    \label{fig:llr_pam4}
\end{figure}

\begin{table}[!t]
    \centering
    \caption{Simulation parameters and values}
    \begin{threeparttable}
    {\renewcommand{\arraystretch}{1.2}
    \footnotesize
    \begin{tabular}{lcl}
    \toprule
    \textbf{Parameter} & \textbf{Value} & \textbf{[Unit]} \\
    \midrule
    Symbol rate & $R_\mathrm{s}$ & [GBd]\\
    Optical modulation amplitude & $3.0$ & [dBm]\\
    Extinction ratio & $4.5$ & [dB]\\
    Link losses & $3.0$ & [dB]\\
    TIA input-referred noise & $22$ & [pA$/\sqrt{\mathrm{Hz}}$] \\
    Relative intensity noise\hspace{10pt} & $-143$ & [dB$/$Hz]\\
    Noise bandwidth & $R_\mathrm{s}/2$ & [GHz]\\
    Constant $\sigma^2(X):p_0$ & $4.840\cdot10^{-22}\times R_\mathrm{s}$ & [V$^2$]\\
    Constant $\sigma^2(X):p_1$ & $5.012\cdot10^{-15}\times R_\mathrm{s}$ & [$-$]\\
    \bottomrule
    \end{tabular}}
    \end{threeparttable}
    \label{tab:sim_param}
\end{table}

The comparison of the LLR calculations using \eqref{eq:Lk_2} or \eqref{eq:Lk_awgn} is shown in Fig.~\ref{fig:llr_pam4}. Here, the LLRs are calculated for the system of Fig.~\ref{fig:system_model} for PAM-4 with BRGC $[00,01,11,10]$ and system parameters denoted in Table~\ref{tab:sim_param} with symbol rate $R_\mathrm{s}=200$ GBd. The results in Fig.~\ref{fig:llr_pam4} show that the exact $L_k$ exhibit a clear nonlinear behavior. Note that using max-log in \eqref{eq:Lk_2} results in a piecewise quadratic function of $y$ due to the argument $\frac{(y-x)^2}{2\sigma^2(x)}$ in the exponential of \eqref{eq:cPDF}. Furthermore, $L_k$ does not have any symmetry due to the monotonic increase of the noise variance alongside $x$. On the other hand, the AWGN-like $\Lawgn$ behaves closer to a symmetric piecewise linear function of $y$. Furthermore, it is clear that there are noticeable differences between $L_k$ and $\Lawgn$, which is expected to lead to a performance penalty in the post-FEC BER if the mismatched LLRs $\Lawgn$ are used instead of $L_k$ \cite[Sec.~V]{bosco2003soft}.
\section{Zero-crossing Approximation for LLRs} \label{sec:llr}

We next propose a piecewise linear approximation of \eqref{eq:Lk_2} that is simpler to compute and, as will be shown in the next section, does not incur a performance penalty in terms of BER. The LLRs with magnitude close to zero represent the \ac{LRB} which are critical for SD decoding. Therefore, we approximate the LLRs around $L_k=0$, i.e., the so-called zero-crossing approximation (ZCA) used in works such as \cite{tosato2002simplified} and \cite{alvarado2009distribution}. We propose to obtain the ZC points $y^\mathrm{zc}$ for each $L_k$, and then obtain the slopes of $L_k$ evaluated at those points to design the piecewise linear function. To derive the expressions for the ZC points and slopes, we use the max-log approximation \cite{viterbi1998intuitive} to simplify the summations inside the logarithm of \eqref{eq:Lk_2} into single exponential terms that are easier to handle analytically. The complete derivation for these expressions is shown in Appendix~\ref{app:zca}. 

For ease of notation, we now define $\sigma_i\triangleq \sigma((2i-M-1)\Delta)$ for $i=1,2,\dotsc,M$, and introduce the ZCA for both PAM-4 and PAM-8 next.

\begin{table}
    \centering
    \caption{PAM-4 zero-crossing points and slope values}
    \begin{threeparttable}
    {\renewcommand{\arraystretch}{1.4}
    \footnotesize
    \begin{tabular}{cccc}
    \toprule
    LLR $L_k$ & ZC point index & ZC point $y^\mathrm{zc}$ & Slope of $L_k$ at $y^\mathrm{zc}$\\
    \midrule
    $k=1$ & $1$ & $-\Delta\dfrac{\sigma_3-\sigma_2}{\sigma_3 + \sigma_2}$ & $\phantom{+}\dfrac{2\Delta}{\sigma_2\sigma_3}$\\[4pt]
    \midrule
    $k=2$ & $1$ & $-\Delta\dfrac{3\sigma_2+\sigma_1}{\sigma_2 + \sigma_1}$ & $\phantom{+}\dfrac{2\Delta}{\sigma_1\sigma_2}$\\[8pt]
    $k=2$ & $2$ & $\phantom{+}\Delta\dfrac{\sigma_4+3\sigma_3}{\sigma_4 + \sigma_3}$ & $-\dfrac{2\Delta}{\sigma_3\sigma_4}$\\[4pt]
    \bottomrule
    \end{tabular}}
    \end{threeparttable}
    \label{tab:zc_pam4}
\end{table}

\begin{table}
    \centering
    \caption{PAM-8 zero-crossing points and slope values}
    \begin{threeparttable}
    {\renewcommand{\arraystretch}{1.4}
    \footnotesize
    \begin{tabular}{cccc}
    \toprule
    LLR $L_k$ & ZC point index & ZC point $y^\mathrm{zc}$ & Slope of $L_k$ at $y^\mathrm{zc}$\\
    \midrule
    $k=1$ & $1$ & $-\Delta\dfrac{\sigma_5-\sigma_4}{\sigma_5 + \sigma_4}$ & $\phantom{+}\dfrac{2\Delta}{\sigma_4\sigma_5}$\\[4pt]
    \midrule
    $k=2$ & $1$ & $-\Delta\dfrac{5\sigma_3+3\sigma_2}{\sigma_3 + \sigma_2}$ & $\phantom{+}\dfrac{2\Delta}{\sigma_2\sigma_3}$\\[8pt]
    $k=2$ & $2$ & $\phantom{+}\Delta\dfrac{3\sigma_7+5\sigma_6}{\sigma_7 + \sigma_6}$ & $-\dfrac{2\Delta}{\sigma_6\sigma_7}$\\[4pt]
    \midrule
    $k=3$ & $1$ & $-\Delta\dfrac{7\sigma_2+5\sigma_1}{\sigma_2 + \sigma_1}$ & $\phantom{+}\dfrac{2\Delta}{\sigma_1\sigma_2}$\\[8pt]
    $k=3$ & $2$ & $-\Delta\dfrac{3\sigma_4+\sigma_3}{\sigma_4 + \sigma_3}$ & $-\dfrac{2\Delta}{\sigma_3\sigma_4}$\\[8pt]
    $k=3$ & $3$ & $\phantom{+}\Delta\dfrac{\sigma_6+3\sigma_5}{\sigma_6 + \sigma_5}$ & $\phantom{+}\dfrac{2\Delta}{\sigma_5\sigma_6}$\\[8pt]
    $k=3$ & $4$ & $\phantom{+}\Delta\dfrac{5\sigma_8+7\sigma_7}{\sigma_8 + \sigma_7}$ & $-\dfrac{2\Delta}{\sigma_7\sigma_8}$\\[4pt]
    \bottomrule
    \end{tabular}}
    \end{threeparttable}
    \label{tab:zc_pam8}
\end{table}
\textbf{1) PAM-4 ZCA:} From Fig.~\ref{fig:llr_pam4} it can be seen that $L_1$ and $L_2$ have one and two ZC points (denoted by the circle markers), respectively. The ZC points and slopes of $L_1$ and $L_2$ at the ZC points are given in Table~\ref{tab:zc_pam4}. With the values from Table~\ref{tab:zc_pam4}, the PAM-4 ZCA is defined as
\begin{align}
    L_1^\mathrm{zc} &= \frac{2\Delta}{\sigma_2\sigma_3}\left(y+\Delta \frac{\sigma_3-\sigma_2}{\sigma_3+\sigma_2}\right),\label{eq:L1_zc}\\[2mm]
    L_2^\mathrm{zc} &= \begin{cases}
        \dfrac{2\Delta}{\sigma_1\sigma_2}\left(y+\Delta\dfrac{3\sigma_2+\sigma_1}{\sigma_2+\sigma_1}\right), & y \leq \gamma\\[4mm]
        -\dfrac{2\Delta}{\sigma_3\sigma_4}\left(y-\Delta\dfrac{\sigma_4+3\sigma_3}{\sigma_4+\sigma_3}\right), & y > \gamma
    \end{cases}\label{eq:L2_zc}
\end{align}
where $\gamma$ is the intersection point between the two branches of $L_2^\mathrm{zc}$. The expressions \eqref{eq:L1_zc} and \eqref{eq:L2_zc} are represented by the black-dashed lines in Fig.~\ref{fig:llr_pam4}(a) and Fig.~\ref{fig:llr_pam4}(b), respectively. 

\begin{figure}[!t]
    \centering
    \includegraphics[]{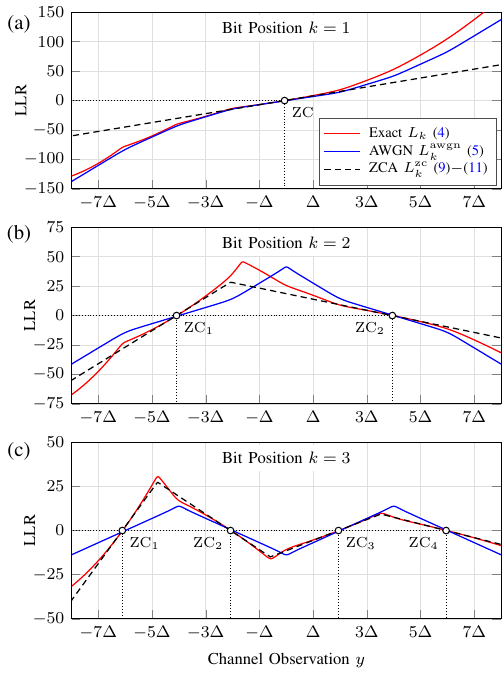}
    \caption{PAM-8 LLRs $L_k$, $\Lawgn$, and $\Lzc$ using parameters in Table~\ref{tab:sim_param} for bit position (a) $k=1$, (b) $k=2$, and (c) $k=3$.}
    \label{fig:llr_pam8}
\end{figure}

\textbf{2) PAM-8 ZCA:} Figure~\ref{fig:llr_pam8} shows the LLRs for the BRGC $[000,001,011,010,110,111,101,100]$. Similarly to the \mbox{PAM-4} case, $L_1$ has one ZC point, while $L_2$ has two. The additional bit level $L_3$ has four ZC points as seen in Fig.~\ref{fig:llr_pam8}(c). The ZC points and slopes of $L_1$, $L_2$, and $L_3$ at the ZC points are given in Table~\ref{tab:zc_pam8}. Hence, the PAM-8 ZCA is designed as
{\allowdisplaybreaks
\begin{align}
    L_1^\mathrm{zc} &= \frac{2\Delta}{\sigma_4\sigma_5}\left(y + \Delta\frac{\sigma_5 - \sigma_4}{\sigma_5 + \sigma_4}\right) \label{eq:L1_zc_pam8} \\[2mm]
    L_2^\mathrm{zc} &= \begin{cases}
        \dfrac{2\Delta}{\sigma_2 \sigma_3}\left(y + \Delta \dfrac{5\sigma_3 + 3\sigma_2}{\sigma_3 + \sigma_2}\right), & y \leq \gamma \\[4mm]
        -\dfrac{2\Delta}{\sigma_6\sigma_7}\left(y - \Delta\dfrac{3\sigma_7 + 5\sigma_6}{\sigma_6 + \sigma_7}\right), & y > \gamma
    \end{cases} \label{eq:L2_zc_pam8} \\[2mm]
    L_3^\mathrm{zc} &= \begin{cases}
        \dfrac{2\Delta}{\sigma_1\sigma_2}\left(y+\Delta\dfrac{7\sigma_2+5\sigma_1}{\sigma_2+\sigma_1} \right), & y\leq \gamma_1\\[4mm]
        -\dfrac{2\Delta}{\sigma_3\sigma_4}\left(y+\Delta\dfrac{3\sigma_4+\sigma_3}{\sigma_4+\sigma_3}\right), & \gamma_1 < y \leq \gamma_2\\[4mm]
        \dfrac{2\Delta}{\sigma_5\sigma_6}\left(y-\Delta\dfrac{\sigma_6+3\sigma_5}{\sigma_6+\sigma_5}\right), & \gamma_2 < y \leq \gamma_3\\[4mm]
        -\dfrac{2\Delta}{\sigma_7\sigma_8}\left(y-\Delta\dfrac{5\sigma_8+7\sigma_7}{\sigma_8+\sigma_7}\right), & y > \gamma_3
    \end{cases} \label{eq:L3_zc_pam8}
\end{align}
}where $\gamma_1<\gamma_2<\gamma_3$ are the intersection points between the adjacent branches of $L_3^\mathrm{zc}$. 

\begin{table}
    \centering
    \caption{Number of operations required to compute all $m$ LLRs}
    \begin{threeparttable}
    {\renewcommand{\arraystretch}{1.5}
    \footnotesize
    \begin{tabular}{ccccc}
    \toprule
    \textbf{Oper.} & \shortstack[c]{Exact\\\eqref{eq:Lk_2}} & \shortstack[c]{AWGN\\\eqref{eq:Lk_awgn}} & \shortstack[c]{Max-log\\\eqref{eq:L_maxlog_awgn}} & \shortstack[c]{ZCA\\\eqref{eq:L1_zc}$-$\eqref{eq:L3_zc_pam8}} \\
    \midrule
    $\sigma^2(x)$ & $M$ & $1$ & $1$ & $M$\\
    $\log(\cdot)$ & $m$ & $m$ & $0$ & $0$\\
    $\exp(\cdot)$ & $mM$ & $mM$ & $0$ & $0$\\
    mul/div & $m(3M-1)$ & $m(2M+1)$ & $m(M+1)$ & $5(M-1)$\\
    add/sub & $m(3M-2)$ & $m(3M-2)$ & $m(M+1)$ & $3(M-1)$\\
    \bottomrule
    \end{tabular}}
    \end{threeparttable}
    \label{tab:complexity}
\end{table}

It can be seen from Fig.~\ref{fig:llr_pam4} and Fig.~\ref{fig:llr_pam8} that the ZCA approximation is designed with lines tangent to $L_k$ at the ZC points. The approximation requires computing the slopes and intercepts according to Table~\ref{tab:zc_pam4} or \ref{tab:zc_pam8}. The channel observation $y$ is then passed through using a linear function, which is simpler in terms of complexity than calculating the value of the exponentials and logarithm of \eqref{eq:Lk_2} or \eqref{eq:Lk_awgn}. The number of math operations required for each LLR calculation is tabulated in Table~\ref{tab:complexity}. From the table, it can be seen that the ZCA is comparable in complexity to the max-log approximation of $\Lawgn$ given by \eqref{eq:L_maxlog_awgn}, as it also omits the operations $\log(\cdot)$ and $\exp(\cdot)$. Given that these transcendental operations are typically precomputed via lookup tables, their omission also results in storage savings. Furthermore, the main trade-off between $\Lawgn$ and the pair $L_k$ and $\Lzc$, is that the former only needs to calculate a single noise variance, while the latter two need to calculate $M$ noise variances, one per symbol. Nonetheless, note from Table~\ref{tab:complexity} that the ZCA LLRs have an asymptotic complexity of $\mathcal{O}(M)$, while exact, AWGN-like, and max-log LLRs have a complexity of $\mathcal{O}(M\log M)$.

\begin{example}[\textit{Post-FEC BER for Extended Hamming Code}] \label{ex:hamming}
As an example of the performance comparison of different LLRs for the SD-FEC decoder, we use the $(128,120)$ extended Hamming code as the inner code \cite{farhood_P802_3dj} in the system of Fig.~\ref{fig:system_model}. We run Monte Carlo simulations with channel observations sampled from \eqref{eq:cPDF}. The symbol rate $R_\mathrm{s}$ is set according to a target bit rate considering the overall code rate of the concatenated FEC, e.g., $R_\mathrm{s}=226$ GBd for PAM-4@400 Gb/s. The rest of simulation parameters are the same as Table~\ref{tab:sim_param}. We sweep the OMA from $-15$ to $10$ dBm, and calculate the BER at every point. For the SD decoder, we use the Chase-II decoding algorithm \cite{chase1972class} with $3$ \ac{LRB} and flipping up to $2$ bits, resulting in a total of $7$ test patterns.
\end{example}

\begin{figure}[!t]
    \centering
    \includegraphics[]{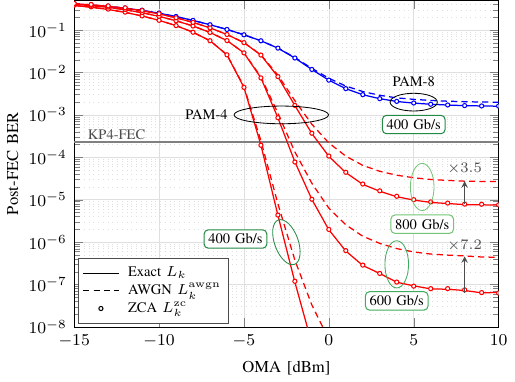}
    \caption{Post-FEC BER against OMA for the $(128,120)$ extended Hamming code with the three different LLR calculations. The group of red curves is the PAM-4 results, while blue is for PAM-8.}
    \label{fig:ber_hamming}
\end{figure}

The post-FEC BER results are shown in Fig.~\ref{fig:ber_hamming}. The results show that the BER converges to an error floor at high OMA due to the presence of RIN. Furthermore, there is a nonnegligible penalty caused by using the mismatched AWGN $\Lawgn$ instead of $L_k$ as predicted in Sec.~\ref{sec:system}-B. For PAM-4, at the KP4-FEC threshold of $2.26\times10^{-4}$ \cite{agrell2018}, the OMA difference between $L_k$ and $\Lawgn$ is around $0.07$, $0.2$, and $0.63$ dB for $400$, $600$, and $800$ Gb/s, respectively. Even more noticeable is the difference in error floors at high OMA, where the error floor with $L_k$ increases by $7.2$ times with $\Lawgn$ for $600$ Gb/s, and by $3.5$ times for $800$ Gb/s. Our low-complexity approximation $\Lzc$ recovers this penalty and matches the exact $L_k$ in terms of post-FEC BER, as can be observed by the markers being on top of the solid curve.\footnote{The results for PAM-4 were presented in \cite{villenas2026ecoc}. In that paper, we neglected one scenario involving the extension bit of the extended Hamming code, which resulted in higher BER values than the ones reported here.} Therefore, we conclude that there is virtually no performance loss when using our piecewise linear approximation $\Lzc$ instead of the quadratic functions resulting from the exact $L_k$, as the soft input for the SD decoder.

Although the error floor of PAM-8 is many orders of magnitude above, the same behavior previously described is exhibited here. In this case, there is a small penalty gap between $L_k$ and $\Lawgn$, and the approximation $\Lzc$ matches the exact $L_k$. The post-FEC BER of PAM-8 is not capable of reaching the KP4-FEC threshold given the amount of RIN present in the channel. This suggests that for these channel conditions, the extended Hamming code is not sufficient to recover from the errors generated by the signal-dependent noise (at this RIN value), and that a code with a stronger error correcting capability is required.

\subsection{Performance with Estimation Errors}
So far, we have seen that the noise variance is critical for calculating the LLRs. Thus, in this section, we discuss the performance of the LLR calculations subject to estimation errors in the noise variances.

\begin{figure}[!t]
    \centering
    \includegraphics[]{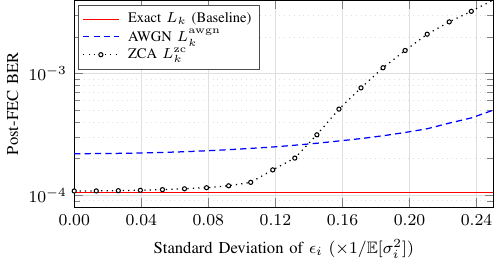}
    \caption{Post-FEC BER performance with noise variance estimation errors. The operating point is PAM-4 at $800$ Gb/s and $\mathrm{OMA}=0$ dBm. The BER of $L_k$ is plotted as a baseline without estimation errors.}
    \label{fig:estimation_error}
\end{figure}

We discuss estimation errors in the single noise variance required for $\Lawgn$, and the $M$ noise variances required for $\Lzc$. We model the noise variance estimation error as an additive Gaussian random variable $\epsilon_i$  with zero-mean and variance $\mathrm{Var}(\epsilon_i)$. To observe the impact of estimation errors, we perform Monte Carlo simulations for the $800$ Gb/s scenario from Fig.~\ref{fig:ber_hamming} at $\mathrm{OMA}=0$ dBm. We transmit a batch of $10^5$ codewords and generate random estimation errors $\epsilon_i$ per batch. The random estimation errors are generated separately for $\Lawgn$ and $\Lzc$. The estimated noise variances are used to calculate $\Lawgn$ and $\Lzc$ for the SD decoder. After decoding, the BER is calculated and averaged over $10^{5}$ different $\epsilon_i$ realizations. The results are shown in Fig.~\ref{fig:estimation_error}, where we show the post-FEC BER against the standard deviation of $\epsilon_i$ normalized by $\mathbb{E}[\sigma_i^2]$. From Fig.~\ref{fig:estimation_error}, it can be observed that when estimation errors are low, $\Lzc$ outperforms $\Lawgn$ for normalized estimation error standard deviations of up to $14\%$. In contrast, if the estimation error is large, using AWGN LLRs achieves a lower BER than ZCA LLRs. For estimation errors below $10\%$, ZCA virtually preserves the BER improvement achieved with ideal variance estimation. Hence, our approximation $\Lzc$ is more susceptible to errors in noise variance estimation, but preserves its improvement over the AWGN-like approach when estimation errors are low. Furthermore, note that for a pilot-based estimation with the same number of pilots, each of the $M$ noise variances required for $\Lzc$ is estimated from approximately $1/M$ of the pilots, resulting in standard deviation of $\epsilon_i$ that is $\sqrt{M}$ times larger than that of the single variance $\overline{\sigma}^2$ required for $\Lawgn$. Hence, we expect $\Lzc$ to require $M$ times as many pilots to achieve the same estimation accuracy.

\subsection{Performance with Residual ISI}

Next, the performance of ZCA is evaluated in a bandwidth-limited scenario. The same system as in Fig.~\ref{fig:system_model} is simulated for different bandwidth constraints. We transmit rectangular pulses at 2 samples per symbol, and the output of the driver is scaled to approximately $0.3V_\pi$ to operate in the linear regime of the MZM and achieve an $\mathrm{ER}\approx 4.5$ dB. The laser power is set to 15 dBm, and the optical power at the MZM output is attenuated to match the target OMA. For fiber propagation, we simulate attenuation and chromatic dispersion at 1310 nm for 500 m of fiber. For the RX DSP we fix the $T/2$-spaced FFE length to 20 taps, and the FFE is trained using the minimum mean square error criterion. After training, the FFE output is downsampled to $1$ sample per symbol. To estimate the noise variance for each PAM symbol, we use a genie-aided approach to calculate $\sigma_i^2=\mathbb{E}[(Y-X)^2|X=x_i]$. On the other hand, for the AWGN-like mismatched metric, the noise estimation is obtained directly as $\overline{\sigma}^2=\mathbb{E}[(Y-X)^2]$. The noise variance estimates are then used to compute the LLRs with the three methods for subsequent decoding with SD. The RIN and thermal noise parameters used are the same as those in Table~\ref{tab:sim_param}.

\begin{figure}[!t]
    \centering
    \includegraphics[]{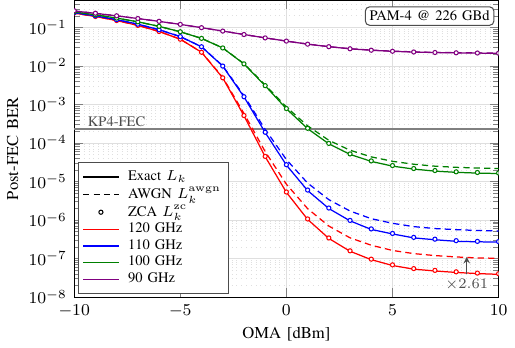}
    \caption{Post-FEC BER against OMA in a bandwidth-limited system. The DAC/ADC bandwidth is set to $\{120,110,100,90\}$ GHz, while the MZM and PD bandwidth is fixed to $120$ GHz for the four different scenarios. The FFE length is fixed for all scenarios.}
    \label{fig:ber_hamming_bwlim}
\end{figure}

The numerical results are shown in Fig.~\ref{fig:ber_hamming_bwlim} for PAM-4 at 226 GBd with four different bandwidth-limited scenarios. The 3-dB bandwidth of the MZM and PD is fixed at 120 GHz, while the 3-dB bandwidth of the DAC and ADC is varied. From the results, it is clear that the BER performance worsens due to the fixed-length FFE not being capable of removing all the ISI as the bandwidth limitation becomes more severe, and also due to noise enhancement from linear equalization. Nonetheless, for the bandwidth-limited scenarios, the results in Fig.~\ref{fig:ber_hamming_bwlim} indicate that, although the penalty from $\Lawgn$ with respect to $L_k$ is much lower compared to Fig.~\ref{fig:ber_hamming}, our approximation $\Lzc$ still matches the BER performance of $L_k$. The performance gap between the pair $L_k,\Lzc$ and $\Lawgn$ starts to decrease as the bandwidth limitation becomes more severe, up to the $90$ GHz case, where there is no difference in performance between the different LLR metrics. This is due to the bandwidth limitation overcoming RIN as the dominant impairment in the system, resulting in poor decoding performance across all LLR calculations.

Lastly, we note that $\sigma_i^2$ will be affected by the frequency response of the electro-optical components and the equalizer. The ZC points will shift accordingly to the dependence on $\sigma_i$. Furthermore, while residual ISI will slightly distort the noise profile at the channel output, we observed that the noise model of \eqref{eq:noise_model} remains a good fit, as seen in the inset (e) of Fig.~\ref{fig:system_model}, by capturing the quadratic dependence with $X$ caused by RIN.

\section{Achievable Rates with SD Decoding} \label{sec:gmi}

In this section, we discuss AIRs with SD decoding considering that the true channel transition probability is given by $f_{Y|X}(y|x)$ in \eqref{eq:cPDF}. The channel under consideration is a \ac{SW} channel, and thus, an AIR for this channel is the \ac{MI} defined as
\begin{equation} \label{eq:mi}
    I(X;Y) =\mathbb{E}_{X,Y}\left[\log_2\frac{f_{Y|X}(Y|X)}{f_Y(Y)}\right]=I(\boldsymbol{B};Y),
\end{equation}
where the right-hand side equality with $I(\boldsymbol{B};Y)$ comes from the fact that $\boldsymbol{B}$ are one-to-one mapped to $X$. However, the system in Fig.~\ref{fig:system_model} considers a \ac{BW} receiver where LLRs are calculated for each bit position, and the bits $B_k$ from $\boldsymbol{B}$ are treated as independent. Therefore, this \ac{BW} receiver can be analyzed via the mismatched decoding framework using the \ac{SW} decoding metric
\begin{equation} \label{eq:q_YgivenB}
    q(\boldsymbol{b},y)\triangleq \prod_{k=1}^m f_{Y|B_k}(y|b).
\end{equation}
An AIR using \eqref{eq:q_YgivenB} is given by the \ac{GMI} which is given by
\begin{align}
    \mathrm{GMI}&\triangleq \max_{s\geq0}\mathbb{E}_{\boldsymbol{B},Y}\left[\log_2\frac{q(\boldsymbol{B},Y)^s}{\sum_{\boldsymbol{b}\in\mathcal{B}^m}P_{\boldsymbol{B}}(\boldsymbol{b})q(\boldsymbol{b},Y)^s}\right] \label{eq:gmi_1}\\
    &=\max_{s\geq0}\sum_{k=1}^m\mathbb{E}_{B_k,Y}\left[\log_2\frac{f_{Y|B_k}(Y|B_k)^s}{\frac{1}{2}\sum_{b\in\mathcal{B}}f_{Y|B_k}(Y|b)^s}\right] \label{eq:gmi_2}\\
    &=\sum_{k=1}^m I(B_k;Y),\label{eq:gmi_3}
\end{align}
where \eqref{eq:gmi_2} comes from using \eqref{eq:q_YgivenB} in \eqref{eq:gmi_1}, and then \eqref{eq:gmi_3} from \cite[Corollary~4.12]{bicmBook} with $s=1$ and from the definition of \ac{MI} in \eqref{eq:mi}. When LLRs are calculated using \eqref{eq:Lk_2} (exact $L_k$), then $I(B_k;Y) = I(B_k;L_k)$ \cite[Th.~4.21]{bicmBook} and the GMI becomes
\begin{equation} \label{eq:gmi_4}
    \mathrm{GMI}=\sum_{k=1}^mI(B_k;L_k).
\end{equation}
In general, the GMI is a lower bound on the MI, and thus $I(X;Y)\geq\mathrm{GMI}$. The equality \eqref{eq:gmi_4} does not hold if LLRs are calculated differently (e.g., $\Lawgn$ or $\Lzc$). In the case of using mismatched LLRs $\tilde{\boldsymbol{L}}$, an AIR is given by \cite[Th.~4.20]{bicmBook}
\begin{equation} \label{eq:gmi_5}
    I^\mathrm{gmi}(\tilde{\boldsymbol{L}})\triangleq m - \min_{s\geq 0}\sum_{k=1}^m \mathbb{E}_{B_k,\tilde{L}_k}\!\!\left[\log_2\!\left(1\!+\!\mathrm{exp}((-1)^{B_k}s\tilde{L}_k)\right)\right]\!\!.
\end{equation}
Note that if the LLRs are set to $\tilde{L}_k=L_k$ in \eqref{eq:gmi_5}, the GMI of \eqref{eq:gmi_4} is recovered, and thus $\mathrm{GMI}=I^\mathrm{gmi}(\boldsymbol{L})$. In summary, we have the following inequalities
\begin{equation}
    I(X;Y)\geq \mathrm{GMI} \geq I^\mathrm{gmi}(\tilde{\boldsymbol{L}}),
\end{equation}
where $I(X;Y)$ can be achieved by an optimal \ac{SW} receiver with perfect knowledge of the channel, GMI can be achieved by a \ac{BW} receiver with matched LLRs, and $I^\mathrm{gmi}(\tilde{\boldsymbol{L}})$ can be achieved by a \ac{BW} receiver with mismatched LLRs \cite{alvarado2016improved}.

\begin{figure}[!t]
    \centering
    \includegraphics[]{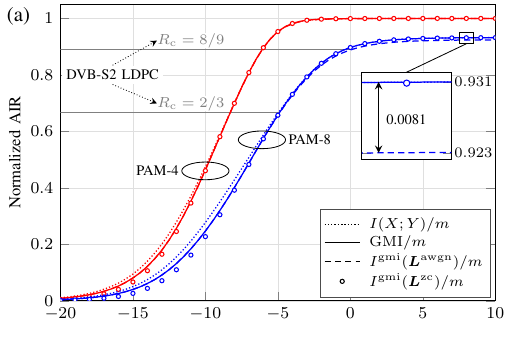}
    \includegraphics[]{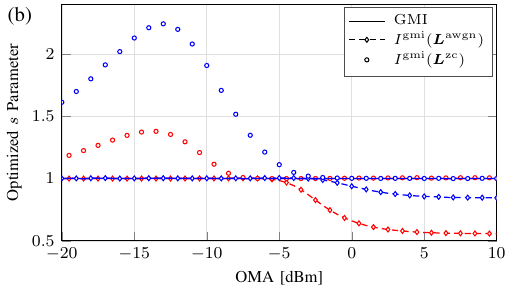}
    \caption{(a) Normalized AIR and (b) optimized $s$ parameter against OMA for the three different LLR calculations using the parameters from Table~\ref{tab:sim_param}. The group of red curves is the PAM-4 results, while blue is for PAM-8.}
    \label{fig:ngmi}
\end{figure}

We now calculate the GMI $I^\mathrm{gmi}(\cdot)$ using \eqref{eq:gmi_5} for the three different LLR calculations $L_k$, $\Lawgn$, and $\Lzc$. The MI $I(X;Y)$ is also included as a reference. The MI and GMI are calculated via Monte Carlo simulation with $R_\mathrm{s}=238.13$ GBd and the same system parameters as in Table~\ref{tab:sim_param}. The results are presented in Fig.~\ref{fig:ngmi}(a) as normalized MI and \ac{NGMI}, i.e., $I(X;Y)/m$ and $I^\mathrm{gmi}(\cdot)/m$, respectively. Additionally, Fig.~\ref{fig:ngmi}(b) presents the results of the $s$ parameter from the optimization in \eqref{eq:gmi_5}.

From Fig.~\ref{fig:ngmi}(a), it can be seen that for PAM-4, $I^\mathrm{gmi}(\boldsymbol{L}^\mathrm{zc})$ achieves a lower rate than GMI in the low OMA regime. Given that the ZCA is based on the max-log approximation, this penalty is consistent with the information loss from the max-log approximation \cite{ivanov2016information}. However, as OMA increases, the markers from the ZCA start to match GMI. Furthermore, the red dashed line of $I^\mathrm{gmi}(\boldsymbol{L}^\mathrm{awgn})$ is indistinguishable from the red solid line of GMI across the entire OMA range. This overlap in GMI does not reflect the results presented in Fig.~\ref{fig:ber_hamming} and Fig.~\ref{fig:ber_hamming_bwlim}, where there are clear differences between the LLR calculations in the error floor region. The reason is that GMI is a predictor of the waterfall behavior of capacity-approaching codes. Therefore, GMI is not a reliable predictor of decoder error floors in the BER for finite-length codes, such as the extended Hamming code.

On the other hand, due to the strong presence of RIN in the channel, the NGMI of PAM-8 saturates around $0.93$. That is, the maximum code rate to achieve a reliable error-free transmission. This also results in a maximum AIR of $664.38$ instead of $714.39$ Gb/s, for $R_\mathrm{s}=238.13$ GBd. Regarding the GMI for the different LLRs, once again it is observed that $I^\mathrm{gmi}(\boldsymbol{L}^\mathrm{zc})$ achieves a lower rate than GMI for low OMA values, and then starts to match GMI as OMA increases towards the saturation region. However, now $I^\mathrm{gmi}(\boldsymbol{L}^\mathrm{awgn})$ only matches GMI for low OMA, and as OMA increases the curve starts to diverge and eventually saturates at a lower value than GMI. The inset of Fig.~\ref{fig:ngmi}(a) shows a clearer picture of this scenario. We observe that the ZCA marker matches well with GMI, while $I^\mathrm{gmi}(\boldsymbol{L}^\mathrm{awgn})$ is below the maximum value by $0.0081$.

The previous conclusions are also supported by the results of the optimization of the $s$ parameter in Fig.~\ref{fig:ngmi}(b). We observe that for $\mathrm{GMI}$ the optimal $s=1$ across the whole OMA range, given that $L_k$ is matched to the true channel \cite[Corollary~4.12]{bicmBook}. For the ZCA $\Lzc$, it can be seen that the parameter $s>1$ for low OMA, exhibiting the mismatch inherited from the max-log approximation. However, the parameter $s$ starts to converge to $1$ as OMA increases, further corroborating that the approximation is a good match for the true channel in this OMA range. The AWGN-like metric $\Lawgn$ shows the opposite behavior to that of $\Lzc$. In the low OMA region, $s=1$, as the channel is dominated by thermal noise. Hence, an AWGN metric is a good match in this regime. However, the parameter shifts to $s<1$ as RIN starts to dominate at higher OMA. Moreover, we observed that scaling the noise variance to $\theta\cdot\overline{\sigma}^2$, with $\theta\in\mathbb{R}^+$, and jointly optimizing $\theta$ with $s$ does not improve the AIR. This is due to the fact that the largest difference in GMI is for high OMA values, where the LLR is approximately $\propto \theta^{-1}\Lawgn$ via \eqref{eq:L_maxlog_awgn}, and thus, the scaling is already captured by the $s$ parameter in the optimization.

\begin{figure}[!t]
    \centering
    \includegraphics[]{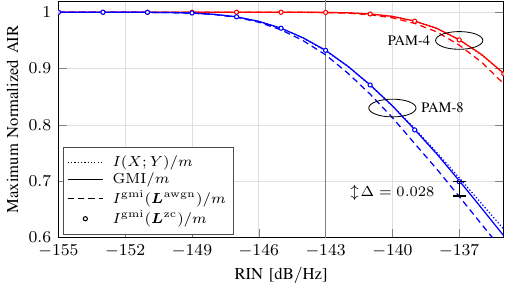}
    \caption{Maximum normalized AIRs against the laser RIN parameter. The AIR values are calculated for $\mathrm{OMA}=10$ dBm.}
    \label{fig:gmi_rin}
\end{figure}

\begin{figure*}[!t]
    \centering
    \includegraphics[]{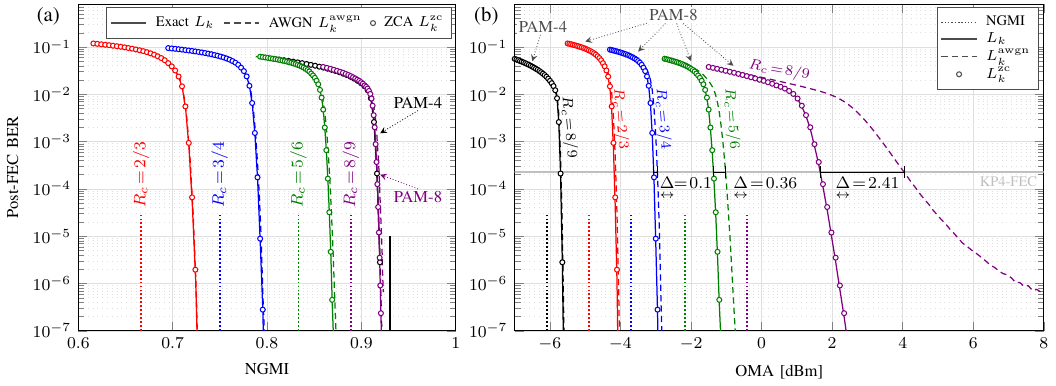}
    \caption{Post-FEC BER of LDPC codes against (a) NGMI and (b) OMA. The colors represent different code rates $R_\mathrm{c}$ for PAM-8, while black is used for PAM-4@$R_\mathrm{c}=8/9$. The curves of PAM-4 and PAM-8 overlap for $R_\mathrm{c}=8/9$ in (a). The dotted lines in (b) represent the OMA value where the NGMI curve is equal to the code rate. The horizontal gaps $\Delta$ are measured in units of dB.}
    \label{fig:ber_ngmi}
\end{figure*}

The dependence of the maximum NGMI value on the laser RIN is shown in Fig.~\ref{fig:gmi_rin}. From the figure it can be seen that for low RIN values $<\!-150$ dB/Hz the NGMI does not saturate and instead reaches a value of $1$. In contrast, as the RIN increases, the maximum NGMI starts to saturate at a lower value. For example, the NGMI of PAM-8 starts to saturate above $-150$ dB/Hz of RIN, while PAM-4 starts saturating around $-142$ dB/Hz, or higher. Furthermore, we observe that the ZCA achieves the same maximum NGMI as the exact LLRs, while the penalty of $I^\mathrm{gmi}(\boldsymbol{L}^\mathrm{awgn})$ also increases along with the RIN value, up to a gap of $0.028$ for PAM-8. 

Next, to evaluate the actual performance of the system based on the NGMI values analyzed, we use LDPC codes as the inner FEC code in the system, as the GMI has been shown to be a good predictor of the BER performance for these types of codes \cite[Sec.~IV-B]{alvarado2015replacing}. In particular, the case of PAM-8 is of most interest, as it is the one that shows the most notable differences in terms of GMI between the LLR calculations and we showed that it requires stronger FEC. The following example tests whether the small NGMI differences observed for PAM-8 due to RIN translate into a significant OMA penalty under practical SD-FEC decoding.

\begin{example}[\textit{Post-FEC BER for LDPC Codes}]
We consider LDPC codes from the second generation digital video broadcasting standard (DVB-S2) \cite{dvbs2}, with code rates $R_\mathrm{c}\in\{2/3,3/4,5/6,8/9\}$. The minimum and maximum code rates are also shown in Fig.~\ref{fig:ngmi} to see where they land in terms of the NGMI. For $R_\mathrm{c}=2/3$, the NGMI is virtually the same for the three LLR calculations. However, for $R_\mathrm{c}=8/9$, while the situation does not change for PAM-4, for PAM-8 this code rate is close to the saturation region of the NGMI, and also where differences between LLR calculations are observable. The code block length is set to $64800$ bits, and the SD-FEC decoder uses the improved min-sum decoding algorithm \cite{chen2005improved}. As in Example~\ref{ex:hamming}, we run Monte Carlo simulations with channel observations sampled from \eqref{eq:cPDF}, using the simulation parameters of Table~\ref{tab:sim_param}. The symbol rate is fixed to $R_\mathrm{s}=238.13$ GBd, i.e., PAM-4@400 Gb/s and PAM-8@600 Gb/s with KP4 plus LDPC with $R_\mathrm{c}=8/9$.
\end{example}

The results for the BER of LDPC codes are shown in Fig.~\ref{fig:ber_ngmi}(a), where the BER from every LLR calculation is plotted against the corresponding NGMI calculated in Fig.~\ref{fig:ngmi}. In Fig.~\ref{fig:ber_ngmi}(a) we use colors to denote the different code rates $R_\mathrm{c}$ for PAM-8, while black is only used for PAM-4 with a code rate of $R_\mathrm{c}=8/9$. From the figure we see that for every $R_\mathrm{c}$, the three different LLR calculations yield essentially identical post-FEC BER vs NGMI performance. In addition, every group of code rates has a steep waterfall, and for $R_\mathrm{c}=8/9$ PAM-4 overlaps with PAM-8. This overlap is consistent with the notion that LDPC codes are universal with respect to GMI \cite{sason2009universal}. Furthermore, note that to achieve a low BER, the required NGMI of the LDPC code with $R_\mathrm{c}=8/9$  is already very close to the PAM-8 NGMI saturation value ($\mathrm{NGMI}=0.931$), indicating that a code with a larger code rate, e.g., $R_\mathrm{c}=9/10$ would not be able to achieve a good BER performance.

The results for the BER of LPDC codes against OMA are shown in Fig.~\ref{fig:ber_ngmi}(b). The vertical dotted lines correspond to the OMA value where the NGMI (from Fig.~\ref{fig:ngmi}) matches the specific code rate. The left-most curves in Fig.~\ref{fig:ber_ngmi}(b) correspond to PAM-4 with $R_\mathrm{c}=8/9$. All three LLR calculations result in very similar BER performance as expected from the NGMI results. On the other hand, we also observe that the PAM-8 curves now have a steep waterfall region, as opposed to the very high error floor of the extended Hamming code of Fig.~\ref{fig:ber_hamming}. However, note that as the code rate increases towards the NGMI saturation value, the rate of change of GMI with respect to OMA decreases. Thus, a larger improvement in OMA is required to lower the BER, given the dependence of the decoding performance on the GMI (as shown in Fig.~\ref{fig:ber_ngmi}(a)). A clear example is the difference between PAM-4 and PAM-8 for a code rate $R_\mathrm{c}=8/9$. The NGMI of PAM-4 has no penalty, resulting in a steep waterfall, while in contrast the NGMI of PAM-8 is very close to the saturation value, and as a result the BER decays more gradually.

Furthermore, the results for $\Lawgn$ (denoted by the dashed curves) exhibit a horizontal gap with respect to the exact $L_k$ (solid lines) in the waterfall region. This gap in OMA (at a BER equal to the KP4-FEC threshold) also becomes larger as the code rate increases. For $R_\mathrm{c}=3/4$ the gap is quantified as $0.1$ dB, while for $R_\mathrm{c}=5/6$ the gap increases noticeably to $0.36$ dB. The code rate $R_\mathrm{c}=8/9$ has the most drastic difference, where now there is a large gap of $2.41$ dB between the two LLR calculation methods. Additionally, below the KP4-FEC threshold, the BER curve of $\Lawgn$ shows a clear tendency to an early error floor, while $L_k$ and $\Lzc$ do not. This results in a great penalty in terms of OMA when using the mismatched LLR $\Lawgn$, even when the difference in NGMI between the LLR calculations is very small $<\!0.0081$. Lastly, we observe that the ZCA $\Lzc$ markers are placed on top of the exact $L_k$, once again resulting in virtually no penalty in terms of BER. Therefore, this is a clear motivation to use our low-complexity approximation $\Lzc$ over the AWGN-like approach, as the penalties in terms of OMA are quite critical between the two methods when the code rate is near the channel NGMI saturation point. This is particularly relevant when the data rate needs to increase and the symbol rate is limited by the devices' bandwidth. In this scenario, either the symbol rate is pushed beyond the device bandwidth, resulting in more complex DSP and equalization stages, or the code rate is increased towards the NGMI saturation value, in which case our approximation offers good performance as has been shown in this section.

\section{Conclusions} \label{sec:conclusions}

This paper studied different LLR calculations for PAM signaling in next-generation RIN-dominated IM-DD systems. We first showed that the exact LLRs for this channel deviate from the typical AWGN channel due to the presence of signal-dependent noise. Next, we showed that although assuming mismatched AWGN statistics on the LLR calculation might result in a lower complexity calculation, there exist significant penalties in the SD-FEC decoding BER performance. Hence, we proposed a novel low-complexity calculation of the channel LLRs based on a piecewise linear approximation around the zero-crossing points of the true LLRs. 

Performance was studied by computing the AIRs of the channel for the different LLR calculation methods. The GMI results show that when the channel is RIN-dominated, the GMI saturates at a lower value than the modulation format entropy, and some small differences are visible with respect to the different LLR calculations. After SD decoding, our results show that there is virtually no performance loss in terms of post-FEC BER, avoiding a BER penalty of up to $7.2$ times observed in the mismatched AWGN case for the $(128,120)$ extended Hamming code. Although this penalty shrinks in the presence of residual ISI, our approximation still outperforms the AWGN mismatch in this scenario. Furthermore, using LDPC codes with code rates near the NGMI saturation point, we showed that small differences in NGMI between the exact and AWGN-like LLR calculations translate into a penalty of up to $2.41$ dB in terms of OMA for the same target BER, and even lead to an early error floor. 

Future work includes extending our approximation to nonstandard PAM modulation formats, e.g., PAM-6 via mapping $5$ bits into a two-dimensional symbol. Additionally, our approximation can be extended to suit constellation shaping. Lastly, experimental validation of the current simulations will also be considered.

\appendices
\section{Zero-crossing Approximation Derivation} \label{app:zca}
We use the max-log approximation \cite[Eq.~(4)$-$(6)]{viterbi1998intuitive} to approximate the terms of the form $\log \sum_j\exp(\varphi_j)\approx\max_j \varphi_j$ in \eqref{eq:Lk_2}. This results in $L_k$ being approximated as
\begin{equation} \label{eq:L_maxlog1}
    L_k\approx \max_{x\in\XkI} \log f_{Y|X}(y|x) - \max_{x\in\XkO}\log f_{Y|X}(y|x).
\end{equation}
Defining $x_b\triangleq \arg\!\max_{x\in\mathcal{X}_k^b} \log f_{Y|X}(y|x)$ for $b\in\{0,1\}$, and using the PDF \eqref{eq:cPDF} in \eqref{eq:L_maxlog1} results in
\begin{align}
    L_k &\approx \log f_{Y|X}(y|x_1) - \log f_{Y|X}(y|x_0) \label{eq:app_step1}\\ 
    &=\frac{(y-x_0)^2}{2\sigma^2(x_0)}-\frac{(y-x_1)^2}{2\sigma^2(x_1)} + \log\frac{\sigma(x_0)}{\sigma(x_1)}.\label{eq:L_maxlog3}
\end{align}
Expanding \eqref{eq:L_maxlog3} and taking the derivative with respect to $y$, we obtain the slope of $L_k$ as
\begin{equation} \label{eq:slope}
    \frac{\mathrm{d} L_k}{\mathrm{d} y}=y\frac{\sigma^2(x_1)-\sigma^2(x_0)}{\sigma^2(x_0)\sigma^2(x_1)}+\frac{x_1\sigma^2(x_0)-x_0\sigma^2(x_1)}{\sigma^2(x_0)\sigma^2(x_1)}.
\end{equation}
Next, using \eqref{eq:L_maxlog3} to find the ZC point $y^\mathrm{zc}$ such that $L_k=0$, results in the solution of $y$ in $f_{Y|X}(y|x_0)=f_{Y|X}(y|x_1)$. This is the same as symbol-wise maximum-likelihood detection. The exact expression for $y^\mathrm{zc}$ is given by \cite[Eq.~(2)]{chagnon2014experimental}. However, it has been shown that a good approximation for uniform constellations is given by \cite[Eq.~(4.6.9)]{agrawal2012fiber}
\begin{equation} \label{eq:zc_point}
    y^\mathrm{zc} \approx \frac{x_1\sigma(x_0)+x_0\sigma(x_1)}{\sigma(x_0)+\sigma(x_1)}.
\end{equation}
Then, evaluating the slope \eqref{eq:slope} at $y=y^\mathrm{zc}$ results in
\begin{equation}
    \kappa \triangleq \frac{\mathrm{d} L_k}{\mathrm{d} y}\Bigg|_{y=y^\mathrm{zc}}=\frac{x_1-x_0}{\sigma(x_0)\sigma(x_1)}.
\end{equation}
Lastly, the LLR ZC approximation $L^\mathrm{zc}$ is constructed as
\begin{equation}
    \begin{split}
        L^\mathrm{zc}&=\kappa(y-y^\mathrm{zc})\\
        &=\frac{x_1-x_0}{\sigma(x_0)\sigma(x_1)}\left(y - \frac{x_1\sigma(x_0)+x_0\sigma(x_1)}{\sigma(x_0)+\sigma(x_1)}\right).
    \end{split}
\end{equation}
Note that $x_0$ and $x_1$ are the symbols that maximize the PDF around $y^\mathrm{zc}$. Furthermore, for a bit position $k\neq1$ in a BRGC, there exists more than one ZC point, and thus, more than one pair of symbols $x_0$ and $x_1$ depending on the value of $y$.

\bibliographystyle{IEEEtran}
\bibliography{references}

\end{document}